\documentclass[11pt,twoside]{article}

\usepackage{asp2014}

\aspSuppressVolSlug
\resetcounters

\begin{document}

\title{Radio source analysis services for the SKA and precursors}

\author{Simone~Riggi,$^1$ Cristobal~Bordiu,$^1$ Daniel~Magro,$^2$ Renato~Sortino,$^3$ Carmelo~Pino,$^1$ Eva~Sciacca,$^1$ Filomena~Bufano,$^1$ Thomas~Cecconello,$^4$ 
Giuseppe~Vizzari,$^4$ Fabio~Vitello,$^1$ and Giuseppe~Tudisco$^1$ }
\affil{$^1$Istituto Nazionale di Astrofisica (INAF), Italy; \email{simone.riggi@inaf.it}}
\affil{$^2$University of Malta, Malta}
\affil{$^3$University of Catania, Italy}
\affil{$^4$University of Milano-Bicocca, Italy}

\paperauthor{Simone~Riggi}{simone.riggi@inaf.it}{}{INAF}{OACT}{Catania}{}{95123}{Italy}
\paperauthor{Cristobal~Bordiu}{cristobal.bordiu@inaf.it}{}{INAF}{OACT}{Catania}{}{95123}{Italy}
\paperauthor{Daniel~Magro}{daniel.magro.15@um.edu.mt}{}{Univeristy of Malta}{Institute of Space Sciences and Astronomy}{Msida}{}{MSD2080}{Country}
\paperauthor{Renato~Sortino}{renato.sortino@inaf.it}{}{UNICT}{Department of Electrical, Electronic and Computer Engineering}{Catania}{}{95123}{Italy}
\paperauthor{Carmelo~Pino}{carmelo.pino@inaf.it}{}{INAF}{OACT}{Catania}{}{95123}{Italy}
\paperauthor{Thomas~Cecconello}{thomas.cecconello@unimib.it}{}{UNIMIB}{Department of Computer Science, Systems and Communications}{Milano}{}{20126}{Italy}
\paperauthor{Giuseppe~Vizzari}{giuseppe.vizzari@unimib.it}{}{UNIMIB}{Department of Computer Science, Systems and Communications}{Milano}{}{20126}{Italy}
\paperauthor{Eva~Sciacca}{eva.sciacca@inaf.it}{}{INAF}{OACT}{Catania}{}{95123}{Italy}
\paperauthor{Fabio~Vitello}{fabio.vitello@inaf.it}{}{INAF}{IRA}{Bologna}{}{40127}{Italy}
\paperauthor{Giuseppe~Tudisco}{giuseppe.tudisco@inaf.it}{}{INAF}{OACT}{Catania}{}{95123}{Italy}
\paperauthor{Filomena~Bufano}{filomena.bufano@inaf.it}{}{INAF}{OACT}{Catania}{}{95123}{Italy}



\begin{abstract}
New developments in data processing and visualization are being made in preparation for upcoming radioastronomical surveys planned with the Square Kilometre Array (SKA) and its precursors. 
A major goal is enabling extraction of science information from the data in a mostly automated way, possibly exploiting the capabilities offered by modern computing infrastructures and technologies. 
In this context, the integration of source analysis algorithms into data visualization tools is expected to significantly improve and speed up the cataloguing process of large area surveys. 
To this aim, the CIRASA (Collaborative and Integrated platform for Radio Astronomical Source Analysis) project was recently started to develop and integrate a set of services for source extraction, classification and analysis into the ViaLactea visual analytic platform and knowledge base archive. 
In this contribution, we will present the project objectives and tools that have been developed, interfaced and deployed so far on the prototype European Open Science Cloud (EOSC) infrastructure provided
by the H2020 NEANIAS project.
\end{abstract}





\articlefigure[width=.7\textwidth]{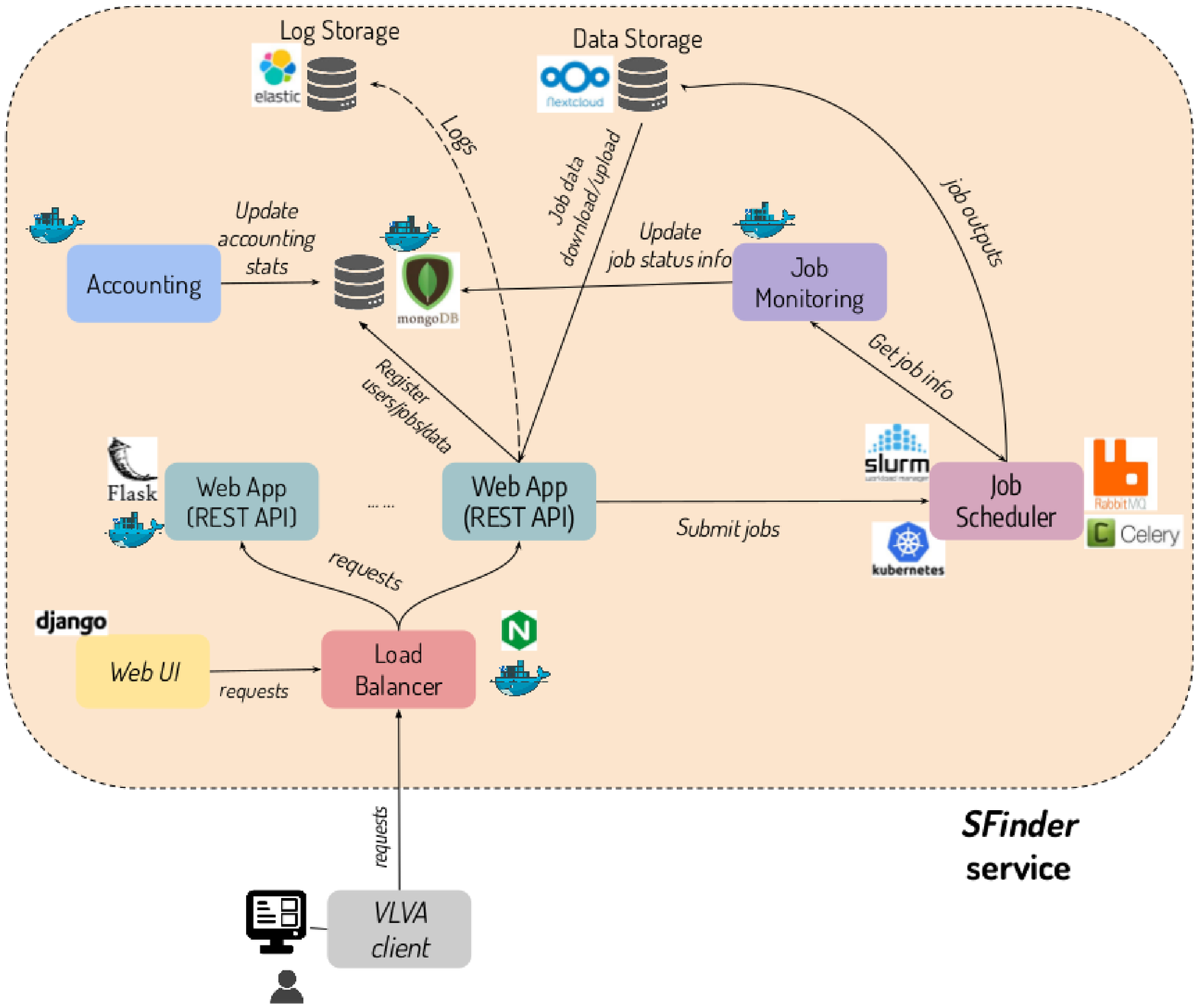}{fig1}{High-level architecture of CIRASA source finding services.}

\section{Introduction}
SKA will make it possible to survey the radio sky with unprecedented level of details, paving the way for
breakthrough discoveries in multiple areas. While the SKA has entered the construction phase, its precursor telescopes are delivering first scientific results. 
The volume of the data are already requiring considerable efforts and new software developments to extract science information 
in an efficient and mostly automated way, anticipating the major challenges expected in SKA Regional Centers (SRCs). 
The availability of distributed services for Advanced Data Products (ADPs) generation from Observatory or Project level data is indeed identified as a key element for delivering SKA science with the SRC. 
SRC Working Groups (WGs) are therefore currently defining requirements and science cases in collaboration with SKA Key Science Project (KSP) users to investigate
ADP management challenges in details.


In this context, we started a new project, named CIRASA \citep{Riggi2021}, to tackle selected SRC cases from the perspective 
of Galactic Science users, using observations done with ASKAP and MeerKAT precursors.


\section{The CIRASA project}
The CIRASA project aims to develop and integrate a set of services for source extraction, classification and analysis into the ViaLactea Visual Analytic (VLVA) platform and knowledge base 
services (VLKB), delivering proto-SRC solutions that can scale to larger computing infrastructures. 
New developments to improve the source cataloguing process will focus on: 
reducing the fraction of spurious sources (currently around 20\%) automatically extracted by standard source finders; detecting objects broken up into multiple source islands; 
selecting sources of likely Galactic origin, possibly identifying their Galactic object class; highlighting unexpected or anomalous objects. In this paper we report the current status of source finding services, while the ongoing activities for 
visualization and archiving services are presented elsewhere \citep{X3-018_adassxxxi,X4-008_adassxxxi}.

\section{CIRASA source finding services}
\emph{caesar-rest}\footnote{\url{https://github.com/SKA-INAF/caesar-rest}} is a Flask-based web service, providing a REST API for uploading input radio images (2D), 
performing source extraction runs on them, and retrieving catalogue outputs. The service architecture, sketched in Fig.~\ref{fig1}, 
consists of multiple containerized micro-services. The web application supports user job submission on a 
Kubernetes or Slurm cluster using Docker and Singularity containers, respectively. User input and job data are stored in a MongoDB database. The service is also integrated with core 
services (AAI, logging and accounting) developed within the NEANIAS H2020 project \citep{X3-012_adassxxxi} for the European 
Open Science Cloud (EOSC) infrastructure. Integration with the VLVA client is in progress. 

The service is deployed on a Kubernetes cluster, set up on the Italian GARR OpenStack cloud\footnote{\url{https://cloud.garr.it/}}, and on CIRASA dedicated resources. 
Users can access it from EOSC marketplace portal\footnote{\url{https://marketplace.eosc-portal.eu/}} through a Django-based web UI. 
At present, the service only allows users to perform CAESAR source finder \citep{Riggi2019} runs, but there are plans to shortly integrate other widely used source finders 
(Aegean, CuTEx, SoFiA). Novel ML-based
applications, described below, will be integrated as soon as the development is completed.


\articlefigure{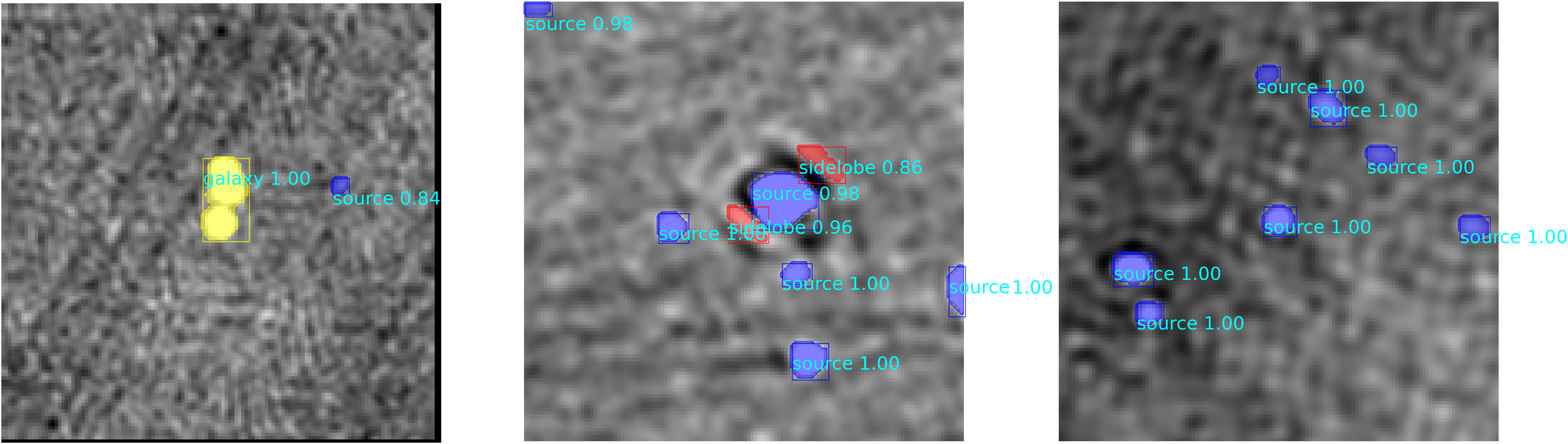}{fig2}{Sample detection results obtained with ASGARD on input images with sidelobes (in red), galaxies (in yellow) and compact sources (in blue).}

\subsection{ASGARD}
ASGARD\footnote{\url{https://github.com/SKA-INAF/mrcnn}} \citep{Magro2021} is a novel source finder based on Mask R-CNN framework, trained to detect three different classes of objects (compact sources, sidelobes and extended radio galaxies) in 2D radio maps. 
Sample results are shown in Fig.~\ref{fig2}. Model performances are rather good for extended galaxies (F$_{1}$>0.9), moderate for compact sources (F$_{1}$>0.75), and poorer for sidelobes (F$_{1}$$\sim$0.3).
ASGARD is already available from \emph{caesar-rest} service interface, but its usage is presently limited to small image cutouts ($<$1000$^{2}$ pixels). To support processing on larger 
maps, a parallel MPI-based version was developed. Further testing activities are needed before full integration in CIRASA.

\articlefigure[width=.6\textwidth]{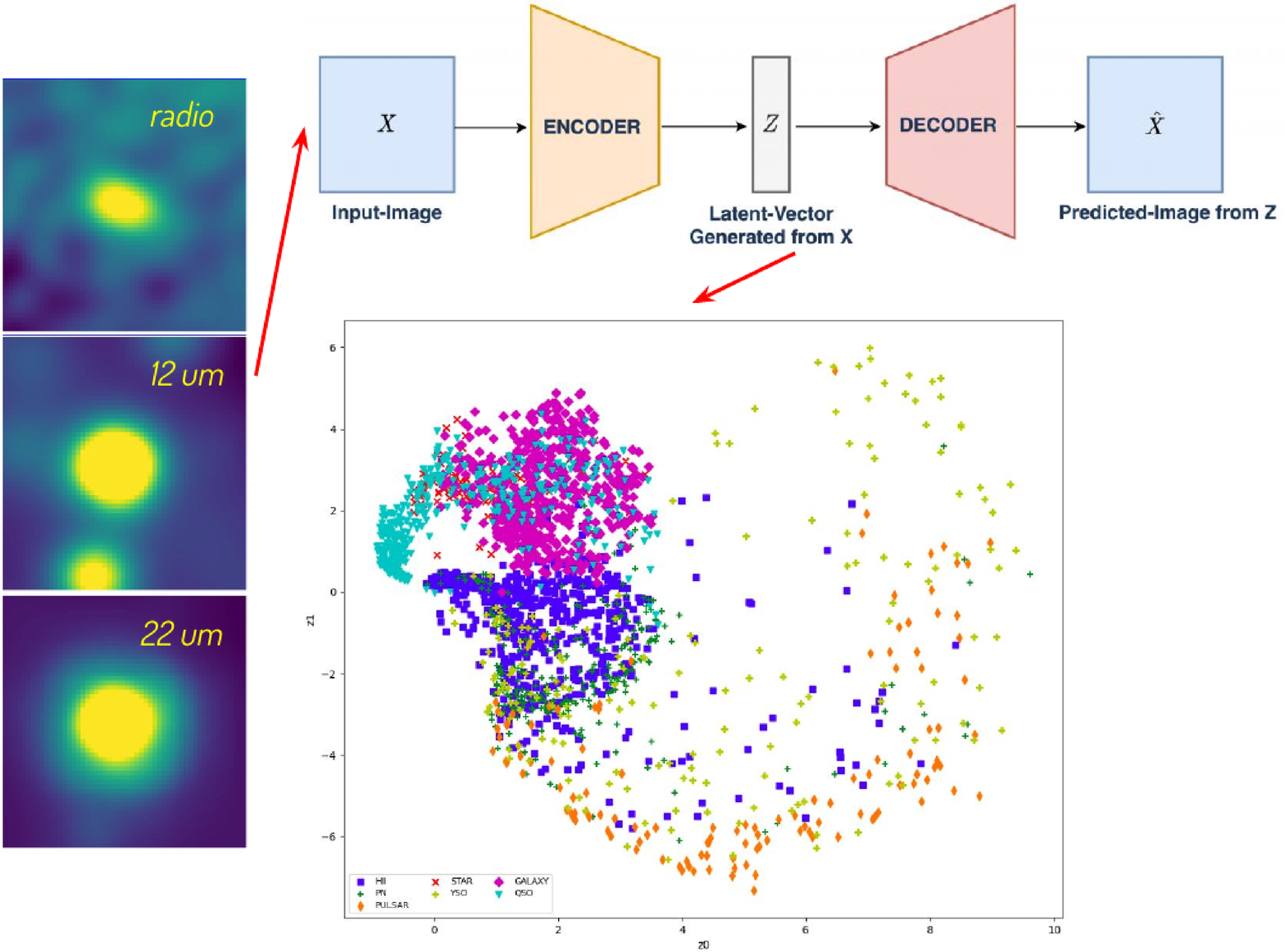}{fig3}{Top: Architecture of autoencoder networks and scatter plot of feature parameters obtained on a sample dataset including different Galactic and extragalactic compact sources.}

\subsection{Compact source classification}
We are also developing new tools for source classification exploiting multiwavelength data. 
In this domain, convolutional autoencoders are a powerful method for extracting feature parameters to produce diagnostic plots or build supervised or unsupervised classifiers. 
In Fig.~\ref{fig3} we report the results obtained with a simple network architecture (2 convolutional layers with 
16 and 8 filters, and a dense layer of size 16) trained to learn a compressed 2D representation of sample input image data (radio, 12 and 22 $\mu$m) for different compact source classes 
(H\textsc{ii} region, PN, pulsar, YSO, extragalactic star, radio galaxy, QSO). As one can see, the trained model already enables potential galactic objects to be selected. Better results are however expected 
to be achieved with additional input infrared bands (e.g. 8 and 70 $\mu$m), feature parameters
(e.g. the radio spectral index), and a more refined network architecture. 
Analysis are ongoing and will be reported in a future work.

\section{Summary and future developments}
In this paper we have presented the source finding services developed so far for the CIRASA project. 
Ongoing activities are focusing on: data preparation for developed applications, source service integration with visualization client and support for other finder applications,  
improvements in deep learning finders. For instance, to overcome some of the limitations found in ASGARD, we are training alternative state-of-the-art models (Tiramisu, YOLOv3, Detectron2, DETR)
on an enlarged dataset. Promising results were found on sidelobe detection, although obtained in some cases at the cost of a net increase in training runtimes. Results will be reported in a forthcoming work.

\acknowledgements This work was financially supported by INAF under the PRIN TEC CIRASA programme and by the European Commission  
under the H2020 grant agreement No. 863448 (NEANIAS).

\bibliography{X1-011}  


\end{document}